\documentclass[final,5p,times,twocolumn]{elsarticle}
\usepackage{amssymb}
\usepackage{bm}
\usepackage{amsmath}
\usepackage{graphicx}
\usepackage{amsfonts}
\usepackage{ulem}
\usepackage{stmaryrd}
\usepackage{tikz}
\usetikzlibrary{shapes}
\usetikzlibrary{calc}
\usepackage{pgfmath}
\usepackage{pgfplots}
\usepackage{float}
\usepackage{subfig}
\usepackage{natbib}
\usepackage{hyperref}
\usepackage{xcolor}

\pgfplotsset{compat=1.18}

\begin{document}
\begin{frontmatter}

    \title{First-principles study on hardening and softening by screw dislocation formation in BCC tantalum alloys}

    \author[1]{Jiajun Feng}

    \author[1]{Kangzhi Zhou}
    \affiliation[1]{
            organization={School of Physics and Electronics, Hunan Normal University},
            city={Changsha},
            postcode={410081},
            province={Hunan,},
            country={China}
        }

    \author[1]{Xiaowei Zhou}
    \author[1]{Xiao Fu}
    \author[1]{Qiuting Luo}

    \author[1,2]{Ziran Liu}
        \ead{zrliu@hunnu.edu.cn}
        \affiliation[2]{
            organization={Key Laboratory of Low-Dimensional Quantum Structures and Quantum Control of Ministry of Education, Key Laboratory for Matter Microstructure and Function of Hunan Province, Department of Physics, Hunan Normal University},
            city={Changsha},
            postcode={410081},
            province={Hunan,},
            country={China}
        }

    \begin{abstract}
      Improving the high-temperature performance and low-temperature plasticity of tantalum (Ta) alloys is a significant scientific challenge. We employed first-principles calculations to study the interaction between screw dislocations and solute atoms in the body-centered cubic (BCC) structure of Ta, with a particular focus on solid solution softening and hardening. We analyzed the impact of various solute elements on the generalized stacking fault energy (GSFE), energy barriers within the single-atom row displacement model, and their interaction with screw dislocations. The results indicate that Hf and Zr, either individually or in combination, exhibit notable solute softening effects in BCC Ta, significantly reducing GSFE, energy barriers, and interaction energies. In contrast, Nb shows relative insensitivity to solute effects, while Mo, W, and Ir demonstrate solute hardening effects. The calculations suggest that the interaction energy between screw dislocations and solute atoms is a reliable indicator for predicting hardening and softening effects. Additionally, we extend these predictions to ternary alloys, demonstrating that the strengthening and softening phenomena in these materials can be explained through the electronic work function at the electronic level.
    \end{abstract}

\end{frontmatter}

\section{Introduction}
Among refractory metals, tantalum (Ta) exhibits exceptional properties, including a high melting point (2996 $^{\circ}$C), a remarkably low ductile-brittle transition temperature (-196 $^{\circ}$C), and excellent plastic processing and forming abilities. Additionally, Ta boasts outstanding corrosion resistance, wear resistance, creep resistance, and high-temperature mechanical properties, making it a valuable material in various high-tech sectors such as aerospace, nuclear industry, metallurgy, and chemical industry, as well as national defense \cite{cardonne1995tantalum, filella2017tantalum, buckman2000new}. Despite these advantageous attributes, it is noteworthy that the ultimate tensile strength of Ta at elevated temperatures falls short of that exhibited by molybdenum (Mo) and tungsten (W). Furthermore, at lower temperatures, the plasticity of Ta diminishes, rendering it susceptible to spallation under intense shock or overload stresses. This limitation restricts its application in specific high-stress and high-impact loading environments \cite{thissell2000experimental}.

Tantalum metal possesses a BCC structure, where the mechanical properties are intricately linked to the behavior of screw dislocations. Screw dislocations represent a fundamental mechanism for plastic deformation in BCC metals \cite{suzuki1991dislocations}, and the exploration of screw dislocations using density-functional theory (DFT) has garnered increasing attention \cite{woodward2001ab, vitek1970core,ventelon2013ab,xu1996atomistic, RAO2023118440, CRPHYS_2021, PRL2023,Clouet2023,nature2022,BIENVENU2022118098,HU2017304}.

The interaction between dislocations and solute atoms is a subject of significant interest, given its pivotal role in a prominent strengthening mechanism known as solid solute hardening (SSH) \cite{klopp1975review,stephens1975role}. Solid solute softening (SSS) can diminish the material's hardness and strength, often leveraged to enhance the malleability and machinability during processing, facilitating easier forming and machining. Extensive experimental studies have delved into the SSS phenomenon in BCC transition metal alloys \cite{sato1973solid, pink1980low,brunner1997effect,okazaki1996solid,nemat2001deformation,HU2017304}, as well as in intermetallic compounds, semiconductors, and ceramic oxide spinels \cite{mitchell1999dislocations}. The addition of alloying elements emerges as an effective strategy to augment the plasticity of BCC metals. Experimental evidence demonstrates that the incorporation of elements such as Re, Os, and Ir into Mo and W can induce SSS \cite{leonhardt1999investigation}. In the case of Ta, experimental findings reveal that the addition of Hf elements diminishes the yield stress of Ta, resulting in low-temperature solid solution softening \cite{gypen1982thermally}. Despite the advancements in understanding the mechanical properties of BCC Ta, comprehensive research on the elements that can effectively strengthen or soften BCC Ta, the mechanisms governing these phenomena, and the precise control of the material's mechanical properties remains lacking.

This study employs three distinct methods to investigate the potential of solutes to induce SSS or SSH phenomena. The calculation of GSFE \cite{shang2014generalized, rice1992dislocation} is utilized to characterize the plasticity of the material, serving as a predictive tool for SSS or SSH. Additionally, the interaction between solute and dislocation \cite{trinkle2005chemistry, ghafarollahi2020theory, rao2023solid} is evaluated through indirect simulation of screw dislocation formation and direct construction of complete screw dislocations. The impact of solute position on the dislocation core is also investigated. Through these methodologies, we analyze the occurrences of SSS and SSH phenomena in both binary and ternary alloys of Ta. The obtained results are compared with previous calculations and experimental findings \cite{GYPEN1979971, gypen1982thermally, mordike1970solid}. Finally, the study delves into the relationship between the electron work function of solute atoms and the phenomena of solid solution softening and hardening. This comprehensive investigation aims to contribute to a more nuanced understanding of the mechanical behavior of BCC Ta, paving the way for informed material design strategies.

\section{Calculation methods}
\subsection{GSFE}
\begin{figure}[h]
	\centering
\includegraphics[width=0.3 \textwidth]{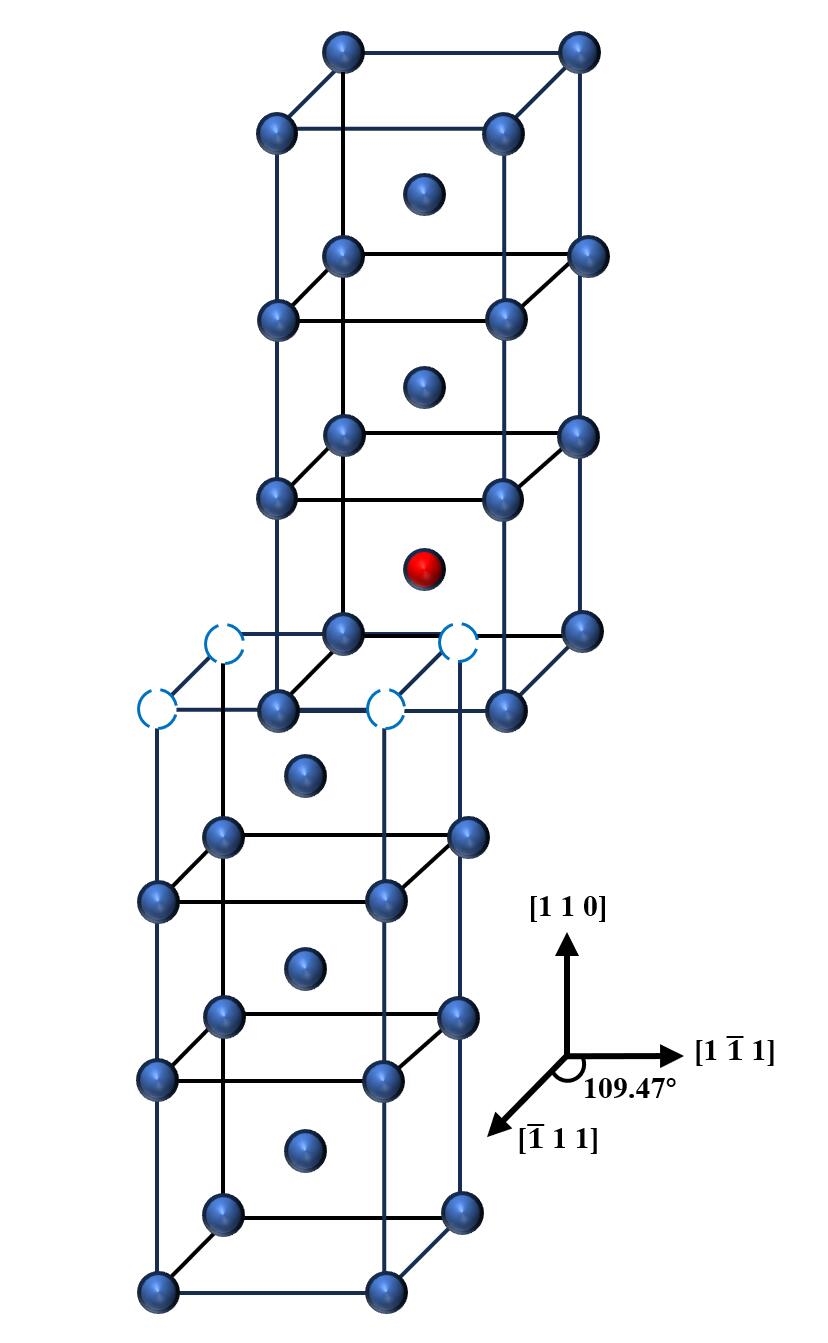}\\
	\caption{The schematic diagram illustrates the geometric structure of the [1$\bar{1}$1]${110}$ slip system. The red ball symbolizes solid solute atoms.}\label{fig1}
\end{figure}

A supercell comprising 12 atoms is initially constructed for a pure BCC Ta bulk. To simulate a binary Ta$_{11}$X$_{1}$ (X=Ir, W, Mo, Nb, Hf, and Zr) bulk, a Ta atom is substituted with an alloy X atom. Stacking fault formation in BCC metals typically occurs along close-packed atomic planes, such as the $\langle111\rangle$ ${110}$ slip system. The supercell, consisting of 12 layers (110) of atomic surfaces, is manipulated to induce stacking faults. To achieve this, the bottom six layers of atoms remain unchanged, while the top six layers of atoms are displaced along the [1$\bar{1}$1] direction, with a slip step of 0.1$\textbf{b}$, where $\textbf{b}$ represents the Burgers vector. This process is illustrated in the schematic diagram of Fig. \ref{fig1}. Following the displacement, atomic relaxation occurs solely in the Z-axis direction, and the total energy of relaxation is computed both before and after the slip.

The magnitude of the GSFE is intricately linked to the plasticity and brittleness of the material \cite{shang2014generalized}. Thus, GSFE serves as a physical quantity to characterize the intrinsic ductility and brittleness of the material. A decrease in GSFE implies a reduced energy requirement for dislocation motion and slip, facilitating plastic deformation and enhancing the material's plasticity.

The GSFE of BCC Ta and Ta$_{11}$X$_{1}$ (X = Ir, W, Mo, Nb, Hf, and Zr) quantifies the energy loss between two adjacent planes during shear deformation in a given slip direction on a designated slip plane. It characterizes the nature of slip and is calculated using the formula \cite{shang2012temperature}:

\begin{equation}
	GSFE=\frac{E-E_{0}}{n S}
\end{equation}
Here, $E$ represents the total energy of the supercell after a certain displacement, $E_{0}$ represents the total energy of the supercell before the displacement, $S$ is the area of the model, and $n$ is the number of stacking faults in the supercell, which is 2 in this case.

\subsection{Single atomic row and screw dislocation models}

We created a 36-atom supercell model to simulate the generation of screw dislocations by displacing a row of atoms along the [111] direction, as described in previous studies \cite{medvedeva2007solid,medvedeva2005solid}. Following the displacement of the atom row, we observed a transformation in the helicity of the three rows of atoms near the simulated dislocation core, aligning with characteristics indicative of an easy core.

To precisely explore the impact of different solutes on solid solution behavior, we constructed a supercell consisting of 135 Ta atoms using the Babel software package \cite{clouet2011dislocation}. This package incorporates the dislocation core field and is modeled within a linear anisotropic elasticity theory with force and dislocation dipoles. Screw dislocations were introduced at 135 Ta atoms with dislocation lines parallel to the [111] direction. In a periodic arrangement, the dislocation core is positioned at the center of gravity of three consecutively neighboring atoms in the [111] direction. The period vectors $\{\mathbf{e}_{1}, \mathbf{e}_{2}, \mathbf{e}_{3}\}$ of the perfect supercell are defined by the elementary vectors $\mathbf{u}_{1}=[\bar{1} \bar{1} 2]$, $\mathbf{u}_{2}=[1 \bar{1} 0]$, $\mathbf{u}_{3}=[111]$ as: $\mathbf{e}_{1}=\frac{5}{2}\mathbf{u}_{1}-\frac{9}{2}\mathbf{u}_{2}$, $\mathbf{e}_{2}=\frac{5}{2}\mathbf{u}_{1}+\frac{9}{2}\mathbf{u}_{2}$, and $\mathbf{e}_{3}=\mathbf{u}_{3}$. The dislocation cores in this supercell are easy cores, which are the most stable type of cores in bcc transition metals.

\subsection{First-principles calculations}

The first-principles calculations are grounded in Density Functional Theory (DFT) \cite{hohenberg1964inhomogeneous} and are executed using the Vienna ab initio simulation package (VASP) \cite{kresse1996efficiency}. For electron-ion interactions, the projector augmented wave (PAW) method \cite{kresse1999ultrasoft} is employed, and the exchange-correlation effects are treated within the generalized gradient approximation of Perdew-Burke-Ernzerhof (GGA-PBE) \cite{perdew2008restoring}. The Monkhorst-Pack framework is utilized for Brillouin zone sampling \cite{pack1977special}, and the obtained results demonstrate that a $k$-point grid of $3 \times 3 \times 15$ is sufficiently accurate. A plane-wave energy cutoff of 450 eV is employed, and the resulting lattice parameter for Ta is $a_{0}$ = 3.30 $\AA$, in excellent agreement with the experimentally measured value of 3.30 $\AA$. The elastic constants are presented in Table \ref{table-e-c}. Throughout the calculation process, both the crystal lattice and atomic positions undergo relaxation to equilibrium. Ionic forces on the atoms are iteratively converged to a value of $10^{-2}$ eV in each relaxation step.

\begin{table}[h]
	\caption{Elastic constants of BCC Ta from previous and this work.}
	\renewcommand{\arraystretch}{0.8}
			\setlength{\tabcolsep}{0.8mm}{ 
				\small
				\begin{tabular}{lcccc}
					\hline
				         $\mathrm{C}_{11}(\mathrm{Gpa})$ & $\mathrm{C}_{12}(\mathrm{Gpa})$ & $\mathrm{C}_{44}(\mathrm{Gpa})$&Methodology&Author \\ \hline
                     \hline
					    273   & 161   & 70& DFT &This work \\
					  271   & 162   & 75 & DFT &  S.I. Rao \cite{rao2023solid}\\
                       271 &164&71&DFT&C.Y. Wu \cite{wu2010effects}\\
					266   & 158   & 87& Exp. &  G.Simmons \cite{simmons1965single}  \\
					  267   & 161   & 82 & Exp.& D. I. Bolef \cite{bolef1961elastic}   \\
				   \hline

			\end{tabular}}\label{table-e-c}
		\end{table}

\section{Results}
\subsection{The GSFE of Ta}
\begin{figure}[h]
	\centering
	\includegraphics[width=0.5 \textwidth]{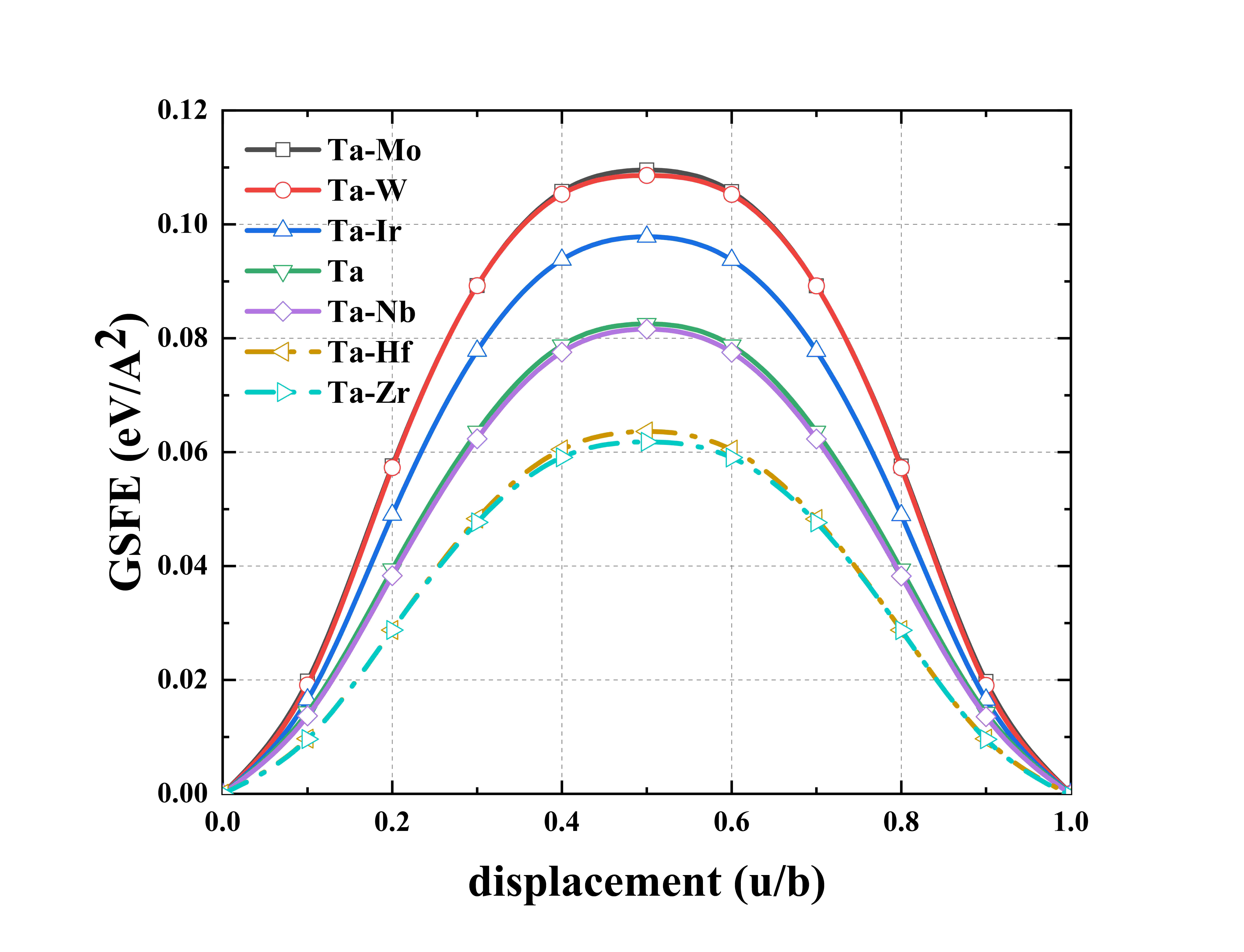}\\
	\caption{GSFE of $<$1$\bar{1}$1$>$ (110) shear for Ta and binary Ta alloys with a slip step of 0.1$\textbf{b}$, where $\textbf{b}$ is the Burgers vector.}\label{fig2}
\end{figure}

In Fig. \ref{fig2}, the GSFE for the $<$1$\bar{1}$1$>$ (110) shear in Ta and binary Ta alloys is shown, considering a slip step of 0.1$\textbf{b}$, where $\textbf{b}$ denotes the Burgers vector. The graph demonstrates a noteworthy rise in GSFE when incorporating solute atoms such as Ir, W, and Mo in comparison to pure Ta, which suggests an elevation in the energy barrier for dislocation motion and slip, thereby rendering plastic deformation more challenging. Conversely, the introduction of solute atoms such as Hf and Zr results in a reduction in GSFE, implying an augmentation in the plasticity of Ta.

\subsection{Single-atom row displacement}
\begin{figure}[h]
	\centering
	\includegraphics[width=0.5 \textwidth]{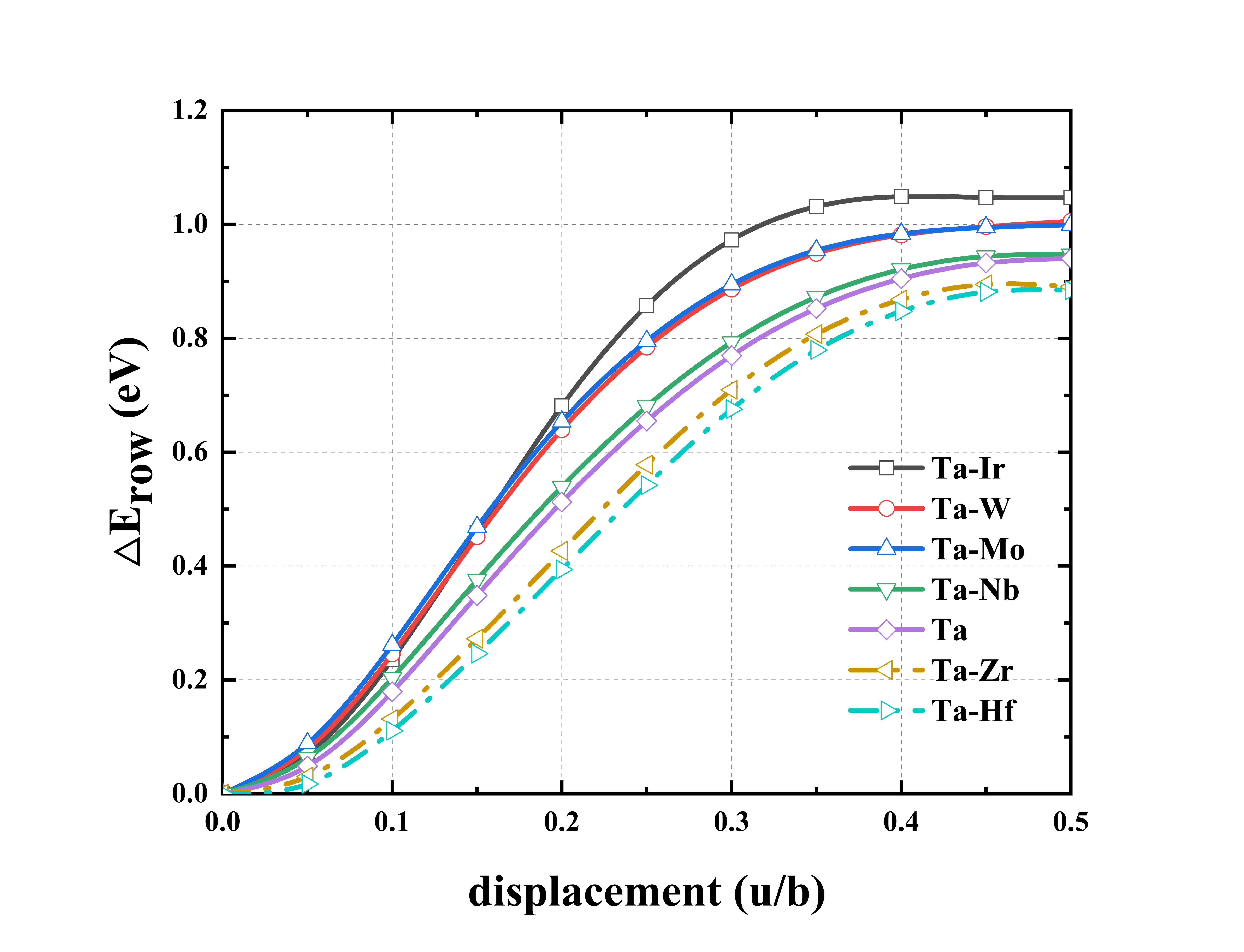}\\
	\caption{ARD energies $\Delta E_{\text {row }}$ resulting from a row of atoms shifting along the [111] direction.}\label{fig3}
\end{figure}

The generation of screw dislocations in BCC metals is closely tied to atomic movement along the [111] direction. To simulate the ease of screw dislocation formation, the slip resistance of single atomic rows along the [111] direction was assessed for pure Ta and Ta alloys containing Ir, W, Mo, Nb, Hf, and Zr solutes. The energy change, denoted as $\Delta E_{\text{row}}$, during the displacement of a single atomic row along the [111] direction was calculated. As shown in Fig. \ref{fig3}, the incorporation of Hf and Zr solutes notably reduces $\Delta E_{\text {row }}$ compared to pure Ta. This reduction enhances kink nucleation and increases dislocation mobility, aligning with observed SSS behavior where Hf induces a lower yield stress in Ta \cite{gypen1982thermally, medvedeva2007solid, medvedeva2005solid}. Conversely, the addition of Ir, W, and Mo increases $\Delta E_{\text {row }}$.

\subsection{Screw dislocations in BCC Ta}
\begin{figure}[h]
	\centering
	\includegraphics[width=0.45 \textwidth]{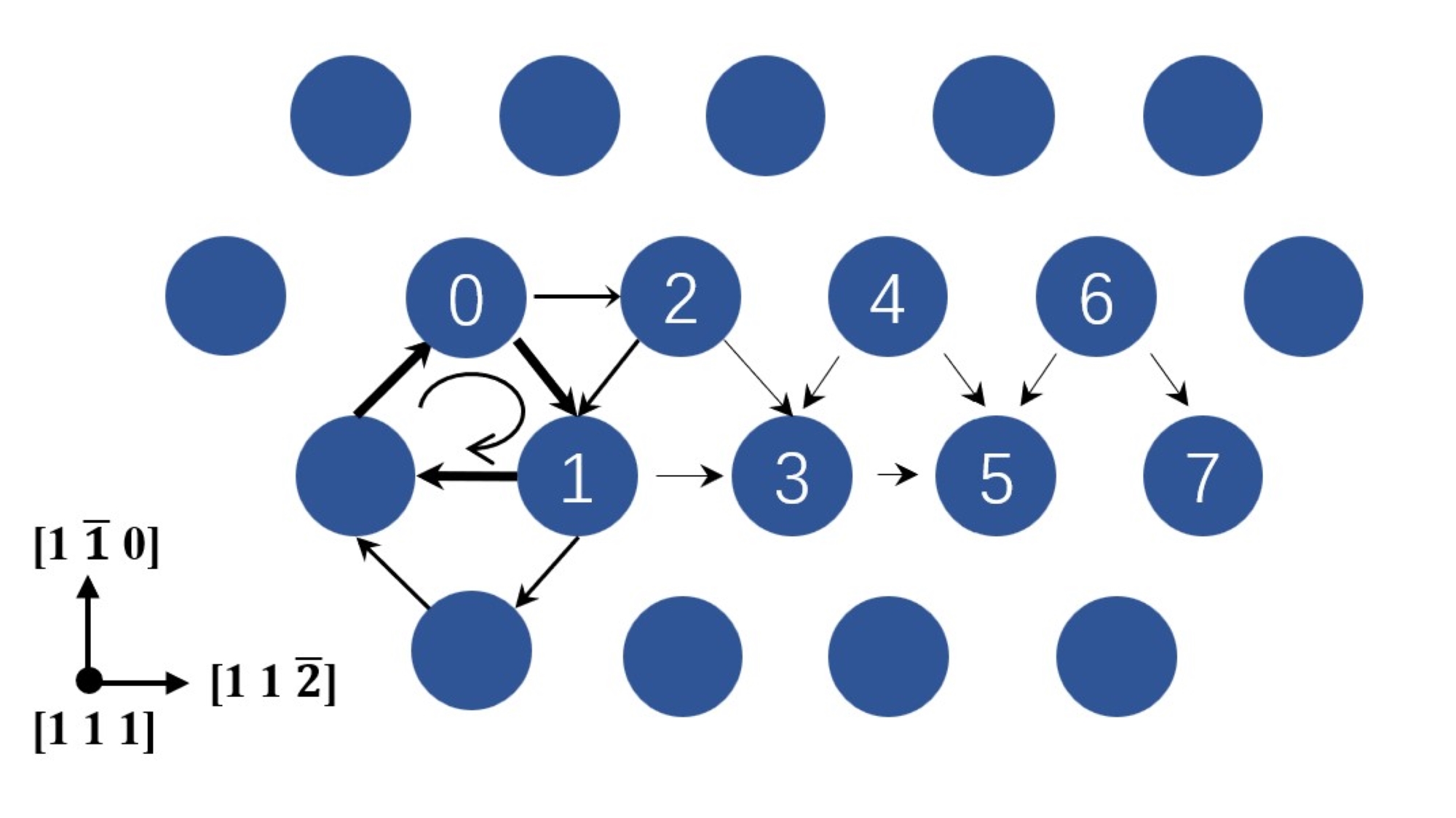}\\
	\caption{Schematic diagram depicting screw dislocations and potential solute atom positions in BCC tantalum (Ta). The arrow curve indicates the location of the screw dislocation core.}\label{fig4}
\end{figure}

\begin{figure}[h]
	\centering
	\includegraphics[width=0.5 \textwidth]{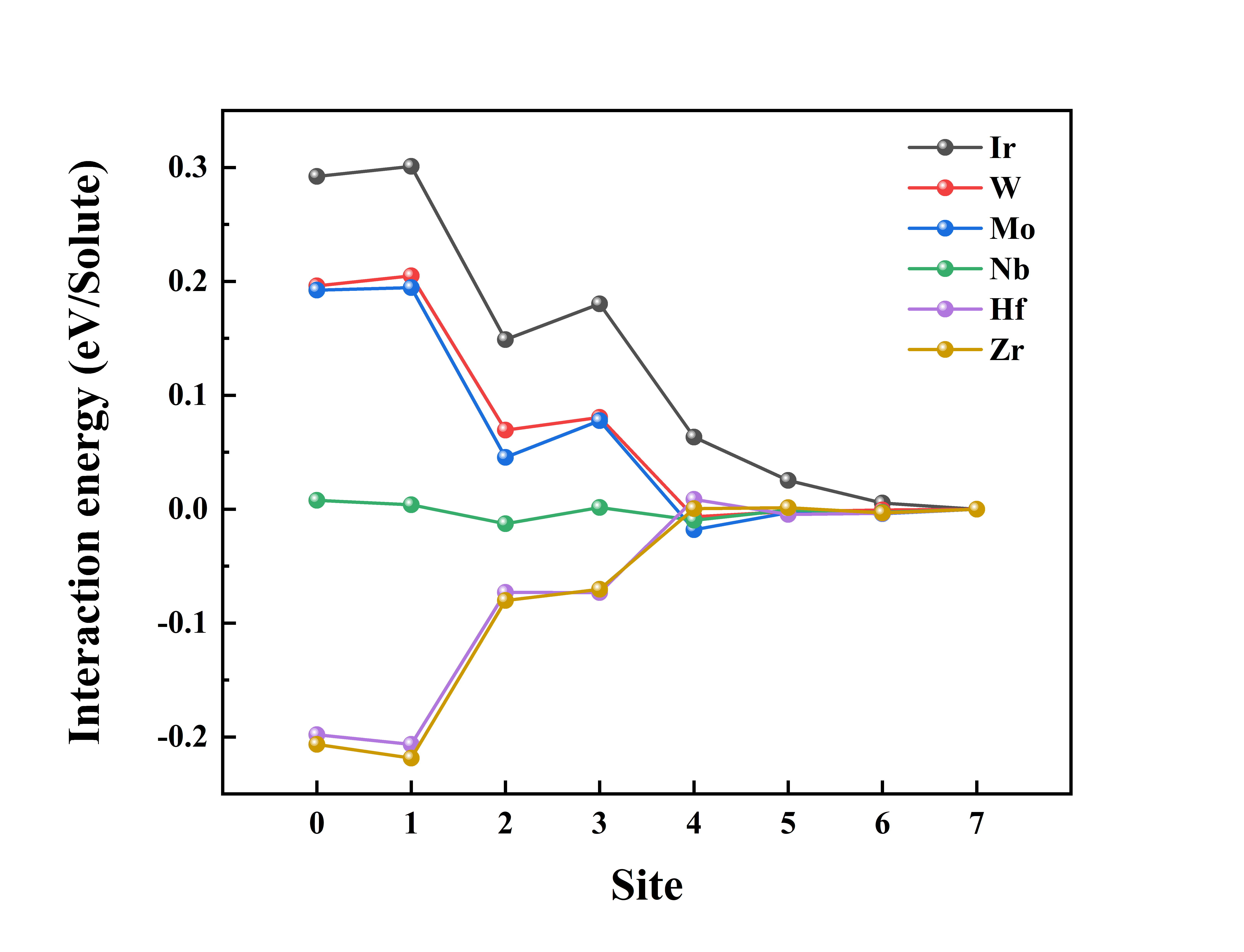}\\
	\caption{Interaction energy variation associated with solutes occupying different sites.}\label{fig5}
\end{figure}

A supercell containing screw dislocation dipoles is constructed to simulate screw dislocations in BCC Ta. Following this, the interaction energy between solute atoms and the dislocation core is calculated, as shown in Fig. \ref{fig4}. Solute atoms are strategically placed at each of the eight positions, labeled 0-7 \cite{tsuru2018first}. The interaction energy is defined as the energy difference between the solute at positions 0-6 and the solute at position 7. Positive and negative interaction energies signify attraction or repulsion between solutes and dislocations. The calculated results are presented in Fig. \ref{fig5}.

The interaction energies at positions 0 and 1 are the largest and nearly equal, as these positions are closest to the dislocation core. As the solute moves away from the core, the interaction energy decreases, approaching zero. Notably, the absolute values of the interaction energies for Hf, Zr, Ir, W, and Mo solutes are substantial when in proximity to the dislocation, while the Nb solute, even in the nearest neighbor of the dislocation core, tends to interact close to zero. This suggests that Nb has a minimal effect on the dislocation compared to the other five solutes. In Fig. \ref{fig5}, energies of Hf and Zr are less than zero, signifying attraction between the solutes and dislocations, whereas Ir, W, and Mo indicate repulsion. These findings align with the results obtained from the 36-atom model, providing consistent insights into the interaction between solutes and dislocation cores.

\section{Discussions}

\subsection{Linear correlation}

\begin{figure}[!htb]
	\centering
	\subfloat[]{
		\includegraphics[scale=0.32]{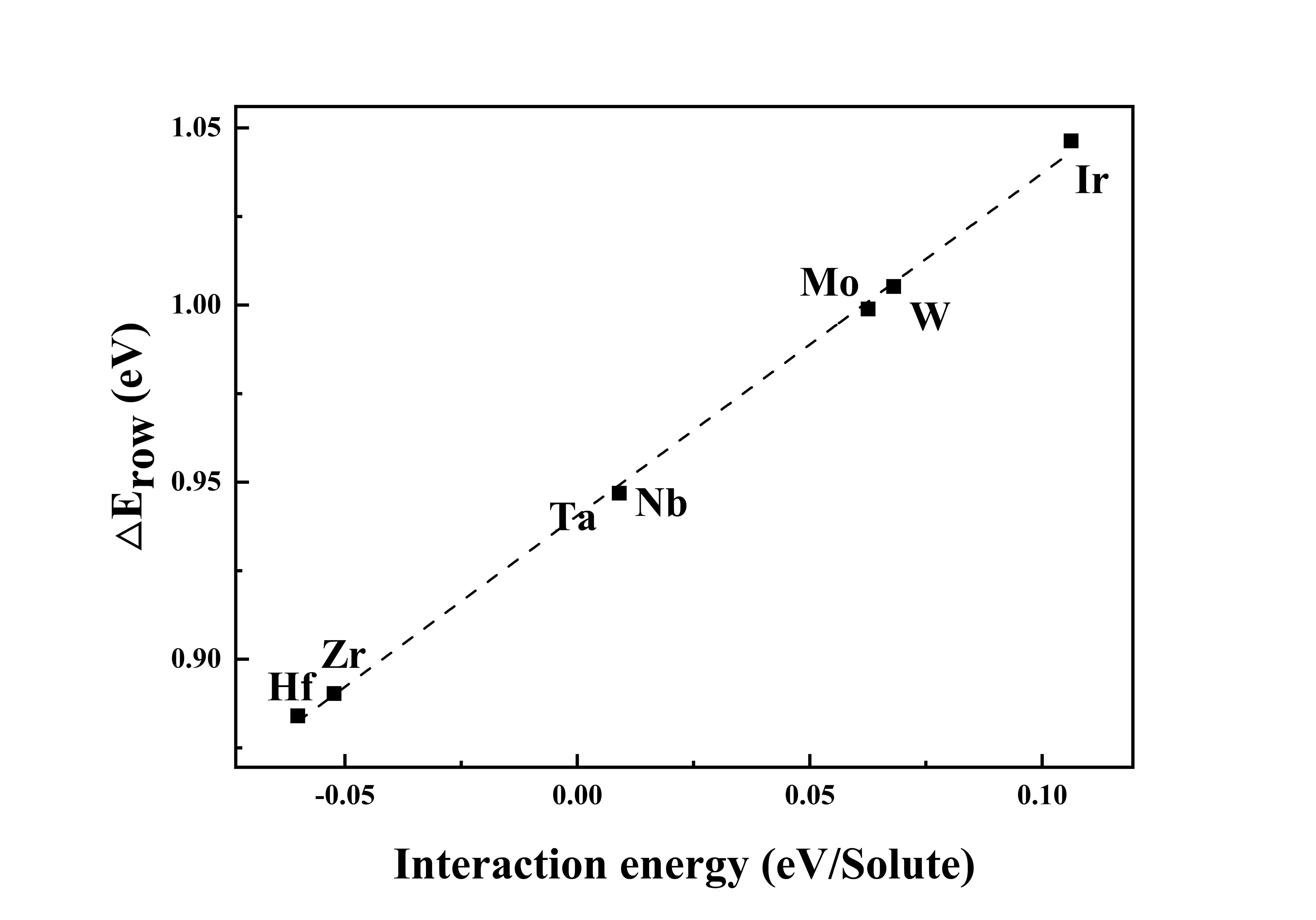}}
	\\
	\subfloat[]{
		\includegraphics[scale=0.30]{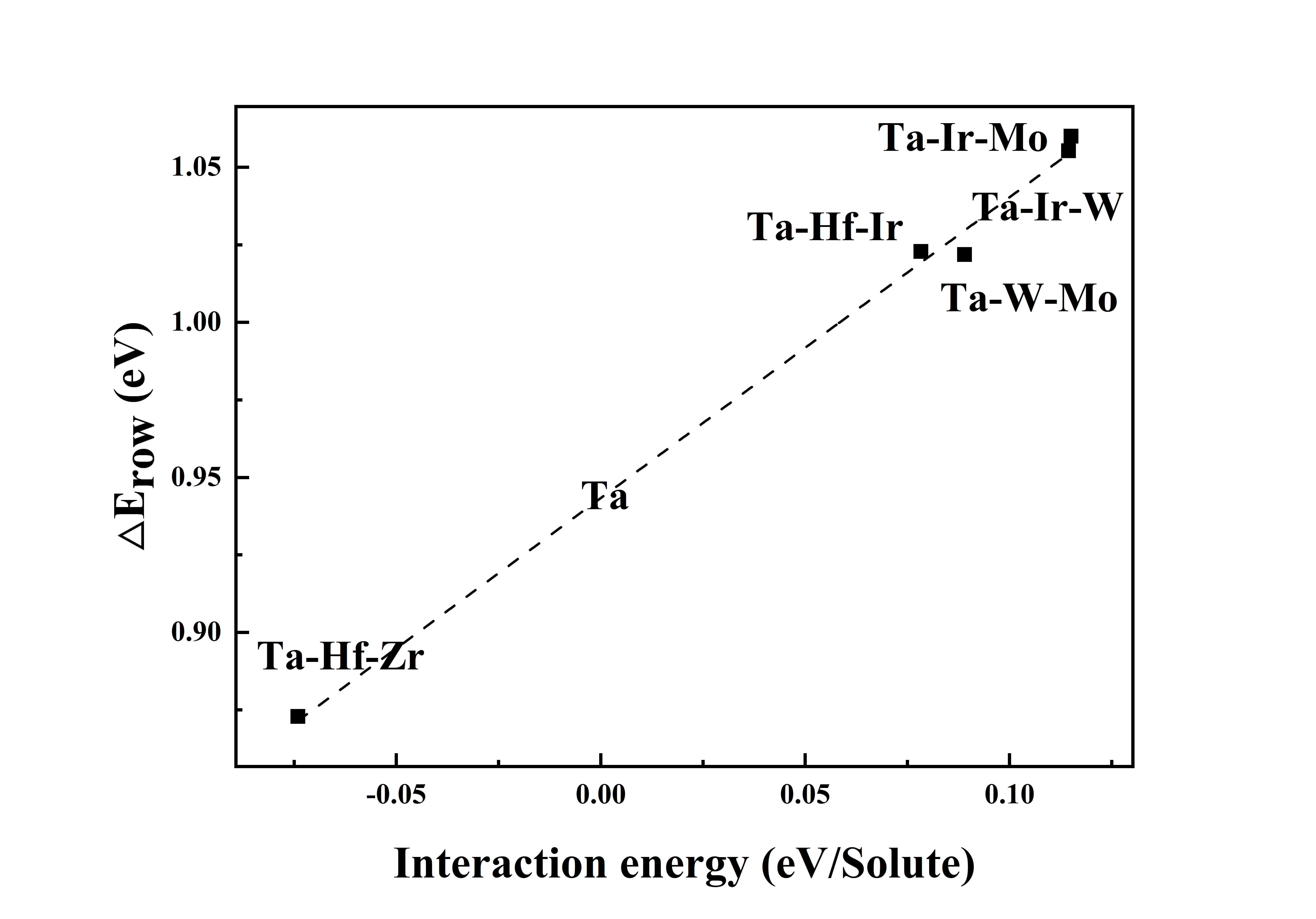}}

	\caption{(a) and (b) are the linear correlation between $\Delta E_{\text {row }}$ and the interaction energy of binary and ternary Ta alloys, respectively.}\label{fig6}
\end{figure}

To gain further insights into how solute atoms impact the mechanical behavior of dislocations, solute-dislocation interaction energy was calculated. This energy reflects the interplay between solutes and dislocations, characterizing the driving force for double kink nucleation. A negative interaction energy suggests an attractive interaction, enhancing dislocation mobility, while a positive interaction energy implies a repulsive interaction. This interaction influences whether the solute addition decreases or increases $\Delta E_{\text {row }}$, contributing to either hardening or softening. The solute-dislocation interaction energies were fitted to ARD energies, revealing a clear linear correlation in Fig. \ref{fig6} (a). This correlation holds significant guidance for binary alloy design. However, the applicability of this rule in ternary alloys with a broader range of applications is explored in the following discussion.

Building upon the analysis of the ternary alloys above, a general deduction can be made: the addition of two solutes together that attract dislocations enhances dislocation mobility, leading to SSS. Conversely, the addition of two solutes that mutually repel dislocations results in SSH. When a solute that attracts dislocations is combined with one that repels the movement of dislocations, the resultant hardening or softening phenomenon depends on which solute has a greater interaction with the dislocations. As shown in Fig. \ref{fig6} (b), a persistent linear relationship exists between atomic displacement energies and solute-dislocation interaction energies in ternary alloys. This relationship enables the prediction of SSH and SSS phenomena based on interaction energies.

\begin{figure}[!htb]
	\centering
	\subfloat[]{
		\includegraphics[scale=0.37]{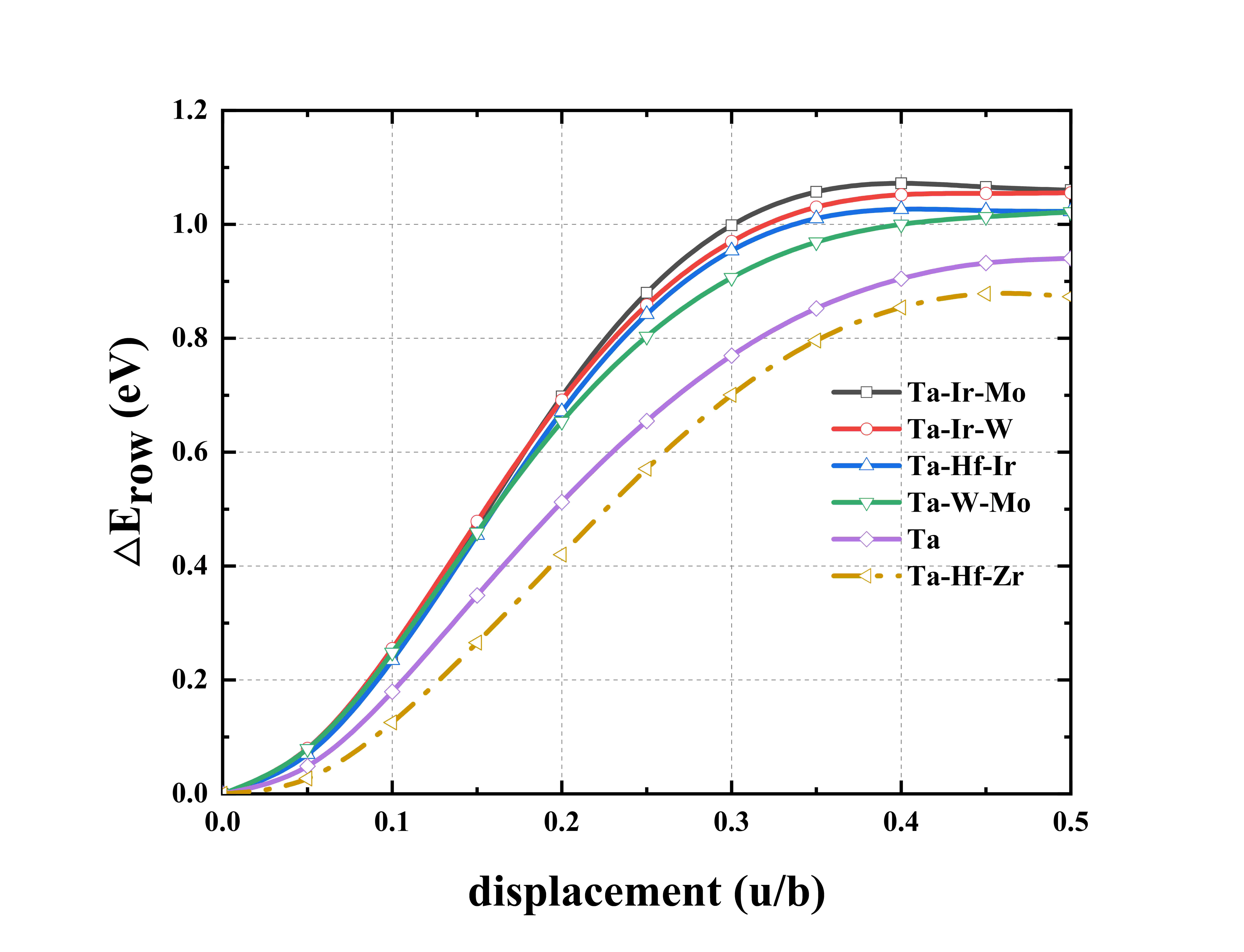}}
	\\
	\subfloat[]{
		\includegraphics[scale=0.35]{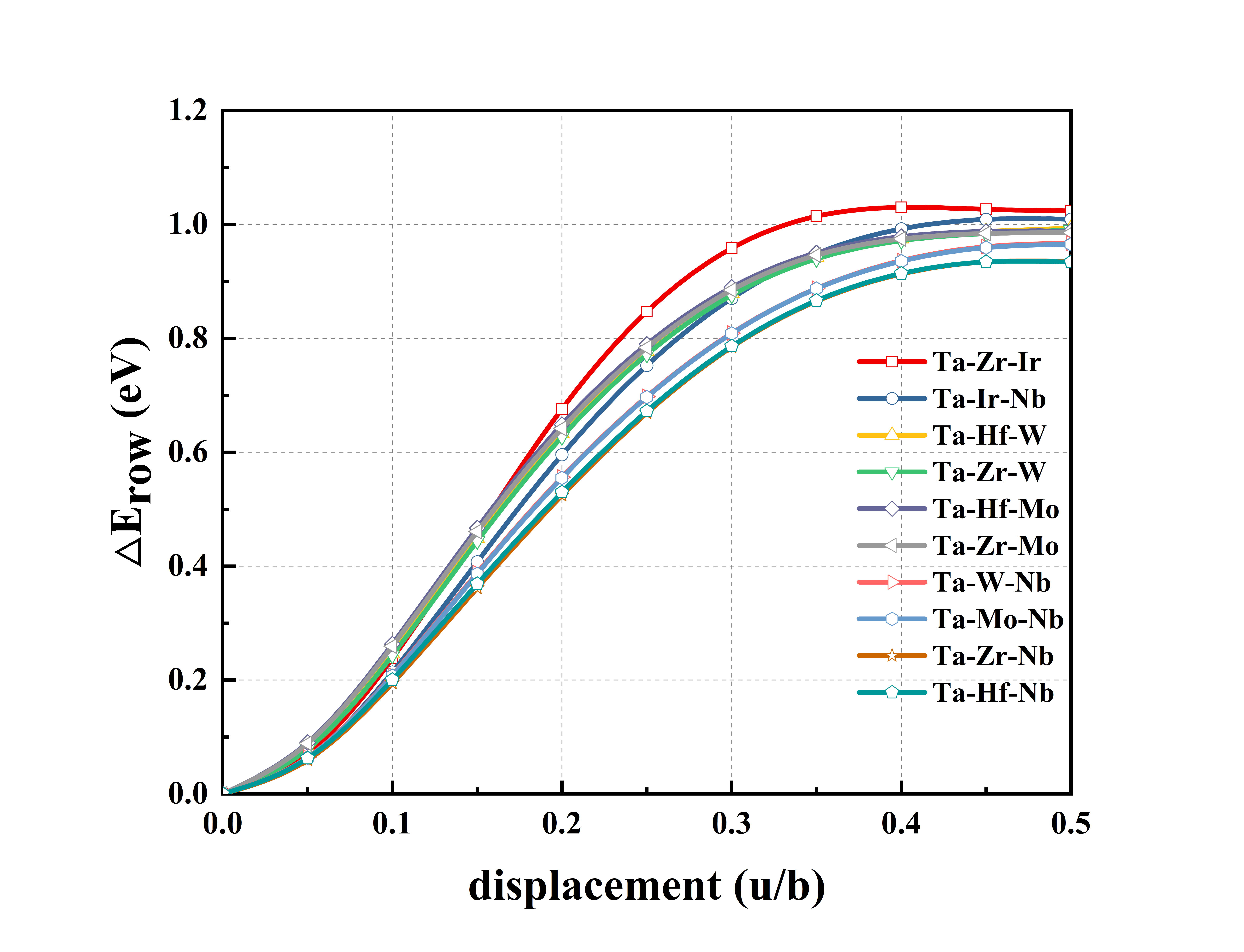}}

	\caption{(a) and (b) show the ARD energies $\Delta E_{\text {row }}$ generated by a row of atoms shifting along the [111] direction in ternary alloys.}\label{fig7}
\end{figure}

Among 15 ternary alloys, five (Ta-Hf-Ir, Ta-Hf-Zr, Ta-Ir-Mo, Ta-Ir-W, and Ta-Mo-W) were selected for illustration. These alloys include solutes previously studied for SSS or SSH behavior in Ta. In Fig. \ref{fig7} (a), the ARD energy of Ta-Hf-Zr is markedly lower than that of pure Ta and also lower than that of Ta-Hf and Ta-Zr, indicating more effective solid solution softening when two softening solutes are added. The ARD energies of Ta-Ir-Mo, Ta-Ir-W, and Ta-Mo-W are notably higher than those of pure Ta and the addition of a single hardening solute, indicating a more pronounced hardening effect. Even with the addition of both a hardening solute and a softening solute to Ta, the Ta-Hf-Ir alloy still exhibits SSH, primarily due to the larger mutual repulsion between Ir and the simulated dislocation core compared to the mutual attraction between Hf and the simulated dislocation core. For the additional ternary alloys illustrated in Fig. \ref{fig7} (b), we have calculated the ARD energies for each. However, predicting the outcomes based on simple rules proves challenging. A more fundamental physical quantity is required, as elaborated upon in the subsequent discussion.

\begin{figure}[h]
	\centering
	\includegraphics[width=0.45 \textwidth]{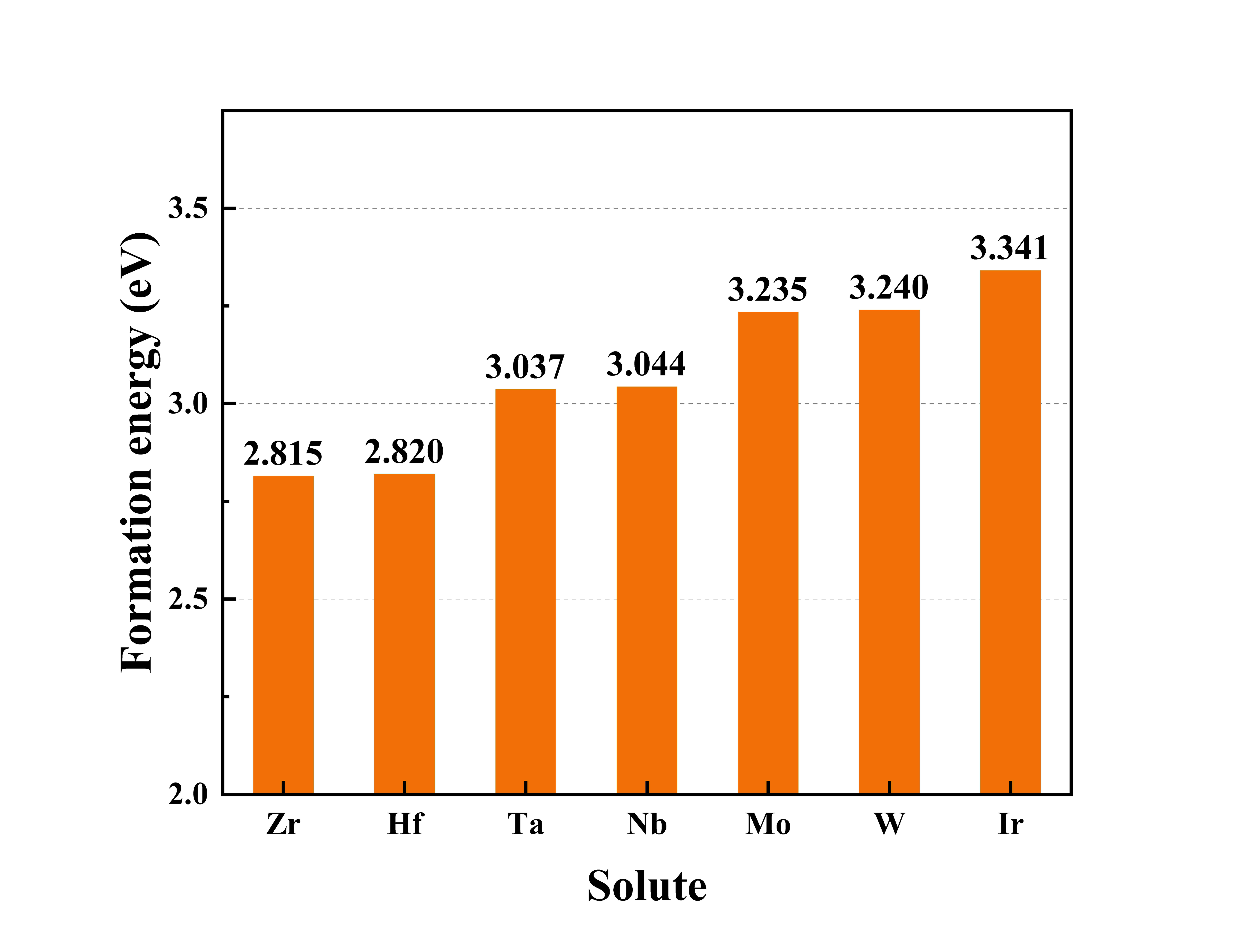}\\
	\caption{Formation energy of screw dislocations following the introduction of various solutes.}\label{fig8}
\end{figure}

To further investigate the influence of different solute atoms on the formation of dislocation cores, we initially calculated the formation energy of dislocations in pure Ta. This is determined by the difference between the energy of two introduced dislocation dipoles and the energy of the perfect supercell. Subsequently, we added various solute atoms and calculated the difference in dislocation formation energy. As shown in Fig. \ref{fig8}, the formation energy significantly decreases after the addition of Hf and Zr compared to pure Ta. Conversely, the addition of Mo, W, and Ir substantially increases the formation energy, while the addition of Nb essentially maintains the formation energy unchanged. A decrease in the formation energy of dislocations suggests that the formation process of dislocations in Ta becomes easier after the addition of solutes, indicating improved plasticity and correlated reductions in strength and hardness of Ta. Therefore, the addition of Hf and Zr may induce a solid solution softening phenomenon in Ta alloys.

\begin{figure}[h]
	\centering
	\subfloat[]{
		\includegraphics[scale=0.35]{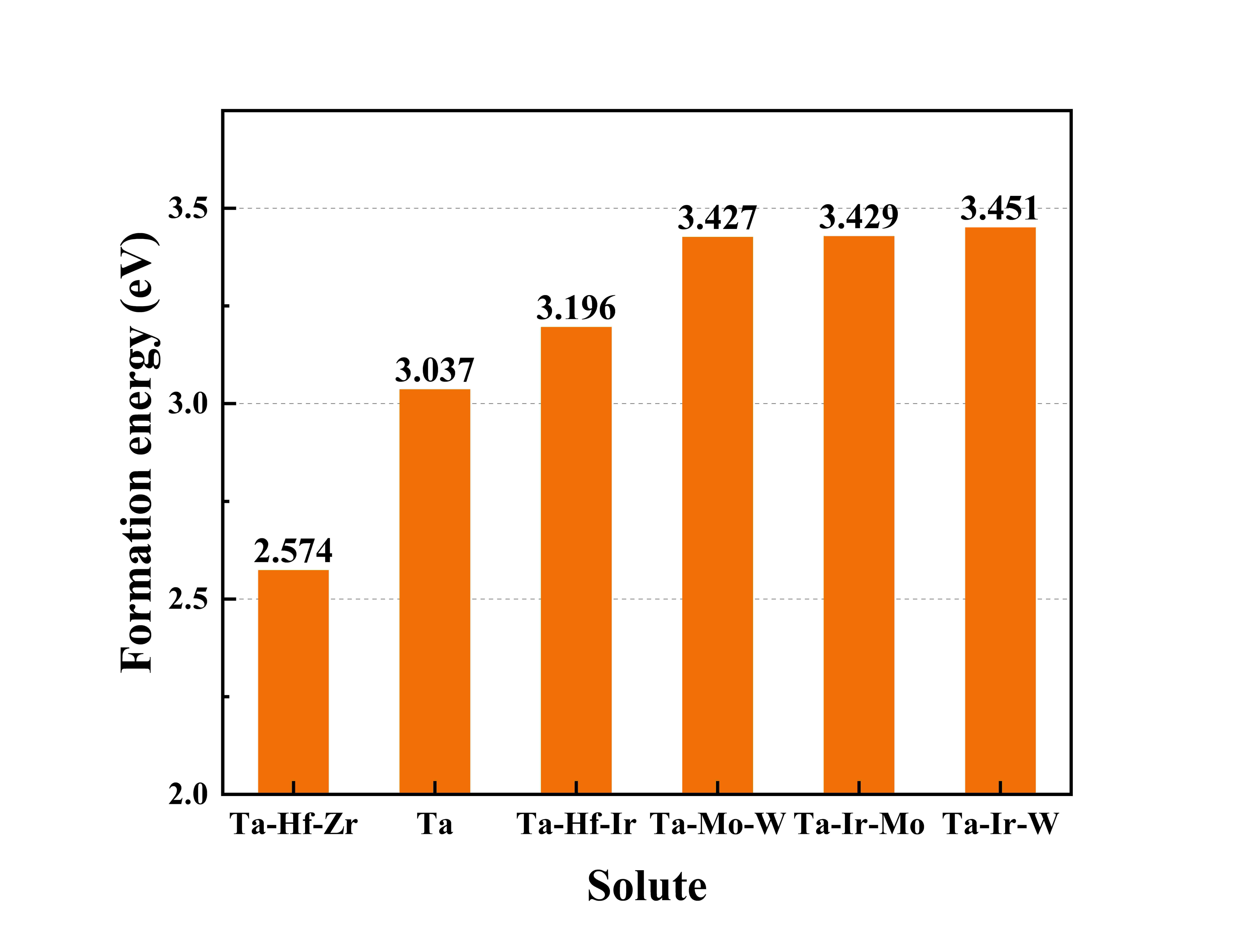}}
	\\
	\subfloat[]{
		\includegraphics[scale=0.34]{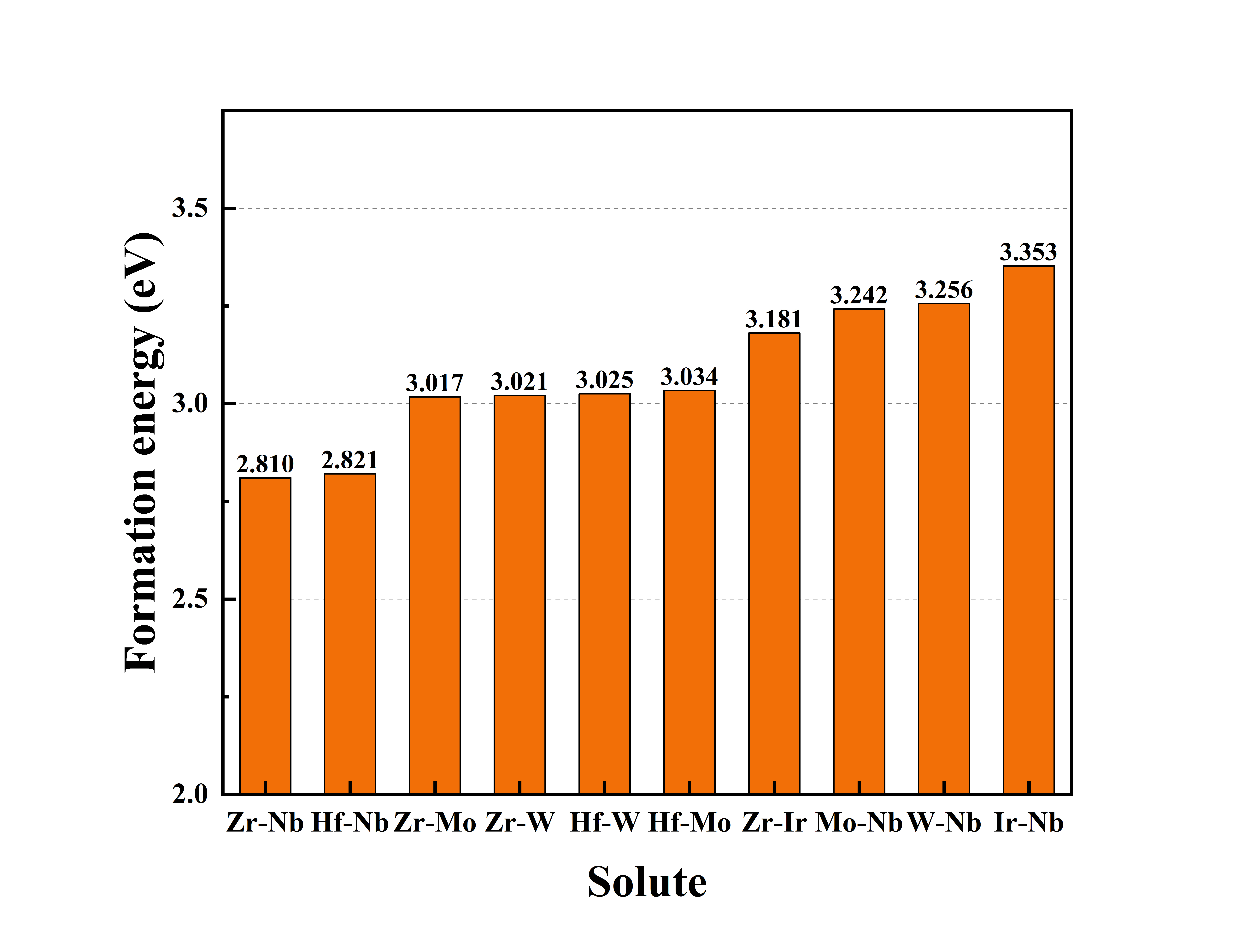}}
	\caption{(a) and (b) Screw dislocation formation energy upon the incorporation of two solutes in Ta alloys.}\label{fig9}
\end{figure}

\subsection{Extend from binary to ternary Ta alloys}

In our extended analysis from binary to ternary alloys, both in single atomic row displacement and screw dislocation dipole models, we found that the SSS or SSH phenomenon in ternary alloys is also related to the interaction energy between solutes and dislocations. Investigating five ternary alloys Ta-Hf-Ir, Ta-Hf-Zr, Ta-Ir-Mo, Ta-Ir-W, and Ta-Mo-W in the model of a complete screw dislocation containing two dislocation dipoles, Fig. \ref{fig9} (a) demonstrates that the dislocation formation energy of dislocation dipoles in Ta-Hf-Zr ternary alloys is lower than that of both Ta and the addition of a single softening solute (Hf or Zr), indicating improved plasticity and lower hardness of Ta-Hf-Zr.

The dislocation formation energies of the ternary alloys with the addition of two hardening solutes Ta-Ir-Mo, Ta-Ir-W, and Ta-Mo-W are all higher than those with the addition of a single hardening solute as well as pure Ta, suggesting a more pronounced hardening effect in these three alloys. Regarding the Ta-Hf-Ir alloy, consistent with the 36-atom model results, the dislocation formation energy of Ta-Hf-Ir is higher than that of Ta due to the significantly larger interaction of Ir with dislocations compared to Hf, leading to the SSH phenomenon. This consistency across the three methods indicates that, compared with pure Ta, the addition of both Hf and Zr can significantly induce SSS, while the addition of W, Mo, and Ir leads to SSH, with almost no change after the addition of Nb.

In other ternary alloys, the strengthening effect persists when two strengthening elements are combined, while a combination of two softening elements leads to a softening effect. The influence of Nb atoms on other solute atoms remains limited. For instance, in Ta-Hf alloys with the addition of Nb, the impact on the formation energy of screw dislocations is not substantial, as shown in Fig. \ref{fig9} (b).

\subsection{Design SSS or SSH by electron work function}

\begin{figure}[h]
	\centering
	\includegraphics[width=0.5 \textwidth]{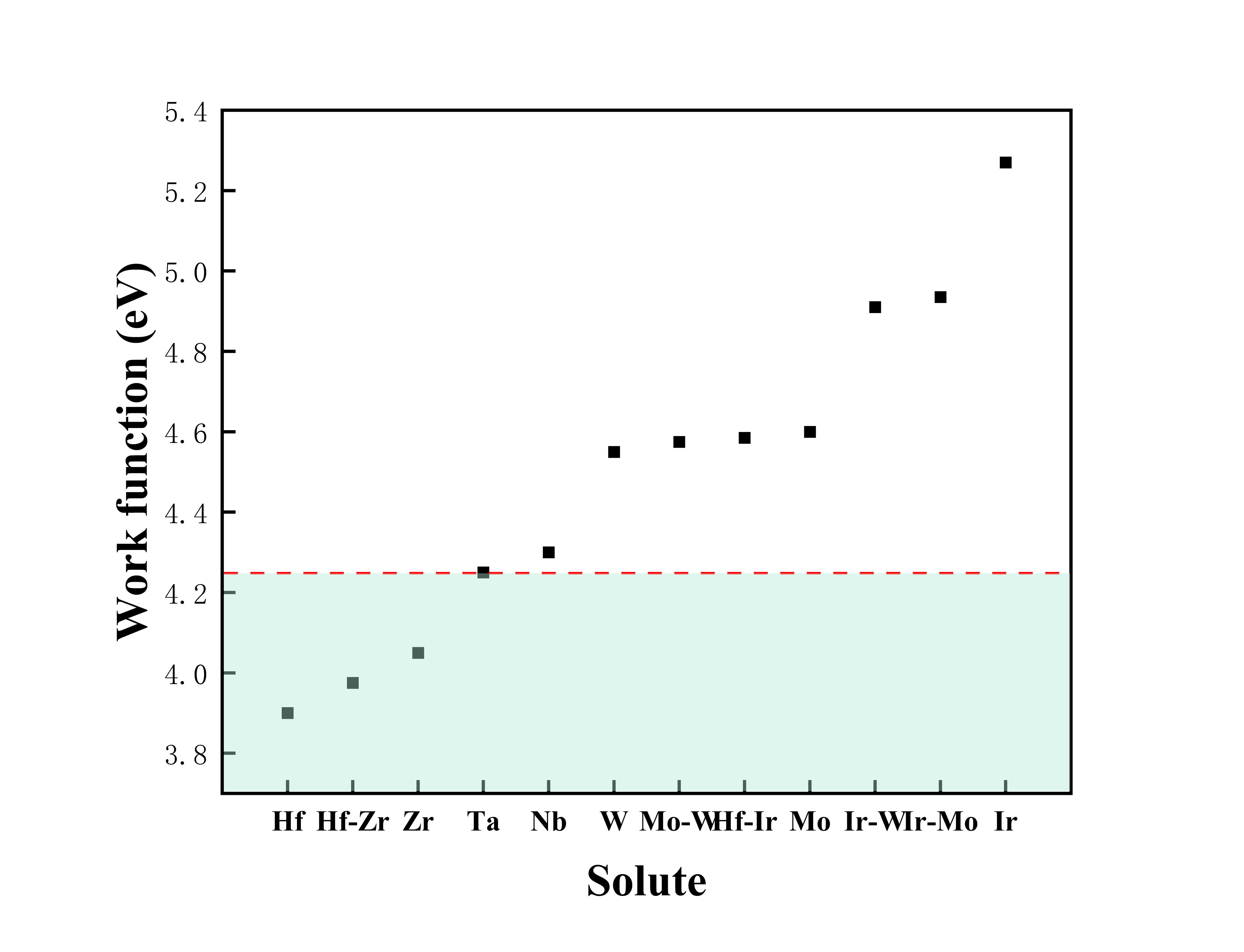}\\
	\caption{Comparison of electron work functions between different solutes and Ta.}\label{fig10}
\end{figure}

Finally, we try to figure out the relationship between the SSS or SSH of BCC Ta alloys with the electron work function (EWF) of the solutes. The EWF represents the energy required to move an electron at the Fermi level from inside a solid to its surface without kinetic energy. This property can be measured using techniques such as the Kelvin probe or ultraviolet photoelectron spectroscopy. While EWF provides insights into the behavior of electrons at the surface, it fundamentally characterizes atomic interactions or the interactions between electrons and nuclei within the bulk of the material. Research indicates that Young's moduli, yield strength, and hardness of metals are inherently correlated with or dependent on EWF \cite{EWF2011}. Moreover, it has been established that EWF serves as a fundamental parameter characterizing electron behavior, not only in pure metals \cite{EWF2011} but also in alloys \cite{LIU2015}. Fig. \ref{fig10} shows that, compared to Ta, the EWF of the solutes that can induce SSS are all smaller than that of Ta, while the solutes that can induce SSH are larger than that of Ta. Nb has almost the same EWF as Ta, indicating that it has a limited effect on Ta. In multi-component alloys, the discussion on the EWF becomes even more meaningful, providing a simpler way to predict whether a solute will lead to SSS or SSH.

\section{Conclusions}
In conclusion, our study systematically investigated the impact of Hf, Zr, Nb, Mo, W, and Ir on the solid solution behavior of BCC Ta using density functional theory and various computational methods. We found that Hf and Zr significantly reduce the GSFE of tantalum, enhancing its plasticity and demonstrating the potential for SSS. The addition of Hf and Zr also lowers the energy barrier for atomic displacement, facilitating dislocation motion and resulting in SSS. By calculating the interaction energy between solutes and simulated dislocation cores, we established a linear relationship that allows the prediction of solute effects on predicting solid solute behavior. Generalizing to ternary alloys, our study predicts potential alloys that may exhibit solid solution softening and hardening in Ta alloys. Finally, our discussion extends to the electron work function of solutes, revealing its correlation with the occurrence of solid solution softening and hardening phenomena. This comprehensive analysis provides a logical framework for understanding the effects of alloying elements on the mechanical properties of BCC tantalum, offering valuable insights for material design.

\section*{Declaration of Competing Interest}
The authors declare that they have no known competing financial interests or personal relationships that could have influenced the work reported in this paper.

\section*{Acknowledgments}
The authors are grateful for financial support from the National Natural Science Foundation of China (51671086) and Hunan Provincial Natural Science Foundation (2022JJ30032).

\bibliographystyle{elsarticle-num-names}
\bibliography{reference.bib}
\end{document}